\documentclass[12pt]{article}
\usepackage[italian,english]{babel}
\usepackage{graphicx}
\usepackage{latexsym}
\usepackage{amsmath,amsfonts,amssymb,amsthm,wasysym}
\def\bseq{\begin{subequation}}  
\def\eseq{\end{subequation}}
\def\bsea{\begin{subeqnarray}}  
\def\esea{\end{subeqnarray}}

\newcommand{\bbox}{\lower.2ex\hbox{$\Box$}}

\newcommand{\beq}{\begin{equation}}
\newcommand{\eeq}{\end{equation}}
\newcommand{\bea}{\begin{eqnarray}}
\newcommand{\eea}{\end{eqnarray}}
\newcommand{\ena}{\end{eqnarray}}

\newcommand {\non}{\nonumber}

\newcommand{\f}{\phi}

\newcommand{\p}{\pi}

\newcommand{\Tr}{{\rm Tr}}

\renewcommand{\[}{\left[}

\newcommand{\be}{\begin{equation}}
\newcommand{\ee}{\end{equation}}
\newcommand{\mc}{\mathcal}
\def\tr{{\rm tr}\ }

\begin{document}
\begin{titlepage}
{\hbox to\hsize{March 2007 \hfill}}
\begin{center}
\vglue .06in
\vskip 40pt

{\Large\bf On meta-stable SQCD with adjoint matter} \\ 
[.13in]
{\Large\bf and gauge mediation} 
\\[.4in]
{\large\bf A. Amariti\footnote{antonio.amariti@mib.infn.it},
L. Girardello\footnote{luciano.girardello@mib.infn.it} and
A. Mariotti\footnote{alberto.mariotti@mib.infn.it}
}
\\[.4in]
{\it Dipartimento di Fisica, Universit\`a degli Studi di
Milano-Bicocca\\ 
and INFN, Sezione di Milano-Bicocca, piazza delle Scienze 3, I 20126 Milano, Italy}
\\[.2in]

\vskip 10pt

{\bf ABSTRACT}\\[.3in]
\end{center}
We briefly review our analysis of a model 
with non supersymmetric vacua in $\mc{N}=1$ gauge
theories with adjoint matter and no R-symmetry.
We show here that this model without any modification
fits into a direct gauge mediation scenario and leads
to massive gauginos.
\vskip 10pt
\vspace{8cm}

\noindent {\em Contribution to the proceedings of the RTN project `Constituents, Fundamental Forces and Symmetries of the Universe' conference in Napoli, October 9 - 13, 2006.}

\end{titlepage}

\section{Introduction}
Following the strategy of the ISS model \cite{Seiberg}
many examples of dynamical supersymmetry breaking in metastable vacua
have been studied 
in supersymmetric $\mc{N}=1$ gauge theories \cite{Uranga,ooguri,Ray,Amariti}
and in string theory \cite{ooguri2,Uranga2,Seiberg2,Ahn,Argurio,Vafa,Tatar,Dudas}.
The approach of ISS relies on theories for which Seiberg-like
dualities exist, i.e. the IR strong dynamics of the electric theory
can be studied perturbatively in
the dual magnetic theory in the range where it is weakly coupled.

In \cite{Amariti} we studied a SQCD-like model with
$SU(N)$ gauge group and adjoint fields with
non trivial superpotential.
Models with adjoint fields
exhibit richer structure. 
In \cite{ooguri,Amariti} it has been necessary to add gauge 
singlet deformations in
order to stabilize the non supersymmetric vacuum.
These models show classical landscape of vacua parametrized by the
adjoint vevs that
can \cite{Amariti} or cannot \cite{ooguri} be wiped out at quantum level.

In the ISS \cite{Seiberg} model the fundamental fields are massive with mass
lower than the natural scale, while in \cite{Uranga,ooguri} the fundamental fields are massless.
In \cite{Amariti} we considered massive fundamental matter but we
will show that our previuos results are valid in the limit of
vanishing quark masses.

The ISS model and its generalizations 
can have phenomenological applications
in connection with gauge mediation 
of dynamical supersymmetry breaking 
to the standard model sector 
\cite{Rattazzi,Fishcler}.
R-symmetry plays here a relevant role since 
a $U(1)$ R-symmetry, even broken to $Z_n$, forbids
a gaugino mass generation.
To obtain a gaugino mass,
deformations can be added to
the superpotential making the R-symmetry trivial,
and this might
require a further careful analysis of its stability.
Quite recently,
meta-stable models 
have been analysed in this direction \cite{Murayama,Dine,ooguri3,Abel,Csaki,Kitano1,Seiberg3}.
In most cases extra terms, breaking R-symmetry,
have been added to known models of dynamical supersymmetry breaking,
leading to gaugino mass at 1 loop at the first or at the
third order in the breaking scale.

Our model \cite{Amariti},
which has meta-stable vacua,
is rather non generic (in the sense of \cite{Nelson}), 
it has no R-symmetry 
and it is suitable for
direct gauge mediation.
We show, indeed, that a 
gaugino mass gets generated at 1 loop at third order
in the breaking parameter.

\section{$\mc{N}=1$ SQCD with adjoint matter}
We consider $\mc{N}=1$ supersymmetric $SU(N_c)$ Yang Mills theory coupled
to $N_f$ massive flavours ($Q^{i}_{\alpha},\tilde Q^{ j \beta}$) 
in the fundamental and
antifundamental representations of the gauge group ($\alpha,\beta=1,\dots N_c$) and 
in the antifundamental and
fundamental representations of the flavour group ($i,j=1,\dots N_f$), respectively.
We also consider a charged chiral massive adjoint superfield $X_{\beta}^{\alpha}$ 
with superpotential%
\footnote{($\Tr$) means tracing on the color indices, while ($\tr)$ on the flavour ones.}
\be
\label{supel}
W_{el}=\frac{g_X}{3} \Tr X^3+\frac{m_{X}}{2} \Tr X^2 +\lambda_X \Tr X 
\ee
where $\lambda_X$ is a Lagrange multiplier 
enforcing
the tracelessness condition $\Tr X=0$.
This theory is asymptotically free in the range $N_f< 2 N_c$ and it admits stable vacua
for $N_f>\frac{N_c}{2}$ \cite{Kutasov2}.
The matching between the microscopic scale ($\Lambda$) and the macroscopic scale ($\tilde \Lambda$)
is
\be
\label{matching}
\Lambda^{2N_c-N_f} \tilde \Lambda^{2 \tilde N-N_f}=\left( \frac{\mu}{g_X} \right)^{2 N_f} \,.
\ee
where the intermediate scale $\mu$ takes into account
the mass dimension of the mesons in the dual description.

We add to the electric potential (\ref{supel}) the gauge singlet
deformations
\be
\label{deltaW}
\Delta W_{el}= \lambda_Q \, \tr Q X \tilde Q  +m_Q \, \tr Q \tilde Q
+ h \, \tr (Q \tilde Q)^2
\ee
The first two terms are standard deformations of the electric
superpotential that
do not spoil the duality 
relations (e.g. the scale matching condition (\ref{matching})) \cite{Kutasov3}.
The last term of (\ref{deltaW}) 
can be thought as originating from a second largely massive adjoint field $Z$ in the 
electric theory with superpotential
\be
W_{Z}=m_Z \Tr Z^2+ \Tr \, Z Q \tilde Q 
\ee
and which has been integrated 
out.
The mass $m_{Z}$ has to be considered larger than $\Lambda_{2A}$, the strong scale of the
electric theory with two adjoint fields.
This procedure leads to the scale matching relation
\be
\label{scaleadj}
\Lambda_{2A}^{N_c-N_f}=\Lambda_{1A}^{2 N_c-N_f} m_Z^{-N_c}
\ee
where $\Lambda_{2A}$ and $\Lambda_{1A}$ are the strong coupling scales 
before and after the integration of the adjoint field $Z$.
The other masses in this theory have to be considered 
much smaller than the strong scale: $\Lambda_{2A}\gg m_Q,m_X$. 
We can suppose that when $h=\frac{1}{m_z}$ is small
the duality relations are still valid.

The dual theory \cite{Kutasov2,Kutasov3,Kutasov1} 
is $SU(2 N_f- N_c \equiv \tilde N)$ with $N_f$ magnetic flavours ($q,\tilde q$), a magnetic
adjoint field $Y$ and two gauge singlets build from electric mesons ($M_1= Q\tilde Q$, $M_2=Q X \tilde Q$),
with magnetic superpotential 
\bea
\label{supmagn}
W_{magn}&=&\frac{\tilde g_Y}{3} \Tr Y^3+\frac{\tilde m_Y}{2} \Tr Y^2+\tilde \lambda_Y \Tr Y 
-\frac{1}{\mu^2}\, \tr \left(\frac{\tilde m_Y}{2} M_1 q \tilde q+\tilde g_Y M_2 q \tilde q 
+ \tilde g_Y M_1 q Y \tilde q \right) \non \\
&&+ \lambda_Q~ \tr M_2  + m_Q~ \tr M_1 + h ~ \tr (M_1)^2 
\eea
For this dual theory the scale matching relation is the same as (\ref{matching})
with $\Lambda \equiv \Lambda_{1A}$ defined in (\ref{scaleadj}).\\
We consider the range where the magnetic theory
is IR free and it admits stable vacua
\be
\label{range}
\frac{N_c}{2}<N_f<\frac{2}{3}N_c \qquad \Rightarrow \qquad 0<2 \tilde N < N_f 
\ee
In this range 
the metric on 
the moduli space is smooth around the origin \cite{Seiberg},
and the 
Kahler potential is regular and can be considered
canonical.

\section{Non supersymmetric meta-stable vacua}
Rescaling the fields and the coupling the superpotential (\ref{supmagn})
is
\bea
W_{magn}&=&\frac{g_Y}{3} \Tr Y^3+\frac{m_Y}{2} \Tr Y^2+\lambda_Y \Tr Y
+ \tr ( h_1 M_1 q \tilde q+h_2 M_2 q \tilde q + h_3 M_1 q Y \tilde q )  \non \\
\label{supmagnbis}
&&-h_1 m_{1}^2 ~ \tr M_1-h_2 m_{2}^2 ~ \tr M_2 + m_3 ~ \tr M_1^2
\eea
Solving the equations of motion we find the 
supersymmetry breaking tree level vacua:
\be
\label{flavours2}
q=\left ( \begin{array}{c}
m_2 e^{\theta}\, \mathbf{1}_{\tilde N}\\
0\\
\end{array}\right) \qquad \tilde q^{T} =\left ( \begin{array}{c}
m_2 e^{-\theta}\, \mathbf{1}_{\tilde N}\\
0\\
\end{array}\right)
\qquad \qquad  
\langle Y \rangle= 
\left( \begin{array}{c  c}
y_1 \mathbf{1}_{n_1}&0\\
0&y_2\mathbf{1}_{n_2} \\
\end{array} \right)
\ee
Where $y_i$ are functions of $n_1$ and $n_2=\tilde N-n_1$ as in \cite{Amariti}.
We choose the vacuum in which
the magnetic gauge group is not broken by
the adjoint field $n_1=0$, which implies $y_2=0$, so $\langle Y \rangle=0$. 
In this case we have 
\be
\label{vuotometa}
\langle M_1 \rangle
=\left( \begin{array}{c  c}
p_1^A&0\\
0&p_1^B\\
\end{array} \right)
\qquad 
\langle M_2 \rangle
=\left( \begin{array}{c  c}
p_2^A&0\\
0&\mc{X}\\
\end{array} \right)
\ee 
where the explicit expressions can be found in \cite{Amariti}.
The two non trivial blocks of the mesons are respectively $\tilde N$ and $N_f-\tilde N$
diagonal squared matrices. 

Supersymmetry is broken at tree level by the rank condition, 
i.e. the $F$ equations of motion of $M_2$ field cannot be all satisfied
\be
0 \neq F_{M_2}=F_{\mathcal{X}}=h_2 m_2^2
\ee
The minimum of
the scalar potential in this tree level vacuum is then different from zero,
and results proportional to $|F_{M_2}|^2$. 
The potential energy of the vacuum 
does not depend on 
$\theta$ and $\mc{X}$;
they are massless fields at tree level,
not protected by any symmetry 
and hence are pseudo-moduli.\\

In \cite{Amariti} the detailed study of the 1-loop quantum corrections to the
effective potential has been performed.
These 
corrections 
depend on the choice of the adjoint vev 
$\langle Y \rangle$: they are minimized
by the choice $\langle Y \rangle=0$.
This is a
true quantum minimum of the scalar potential
where the pseudomoduli get positive mass squared from
the Coleman-Weinberg potential.
The field $\mathcal{X}$ gets a
non trivial vacuum expectation value from the
quantum corrections, $\langle \mathcal{X}\rangle \neq 0$, moving slightly 
the minimum away from the origin.

We report the masses 
for the pseudomoduli 
in the regime of small $\rho=\frac{h_1}{h_2}$ and small $\zeta=\frac{h_1^2 m_1^2}{2 h_2 m_2 m_3}$
($\eta=\frac{2 m_3}{h_2 m_2}$ has been neglected as in \cite{Amariti})
\bea
m_{\mc{X}}^2&=&\frac{\tilde N (N_f -\tilde N)}{8 \p^2}|h_2 m_2|^2 \bigg(|h_2|^2 (\log[4]-1)+|h_1|^2 (\log[4]-2)
\bigg) \non
\\
m_{\tilde \theta}^2&=&\frac{\tilde N (N_f -\tilde N)}{16 \p^2}|h_2 m_2^2|^2 \bigg(|h_2|^2 (\log[4]-1)+\left| 
\frac{h_1^2 m_1^2}{2 m_2 m_3}\right|^2 (\log[4]-\frac{5}{3})+|h_1|^2(2\log[4]-3)
\bigg) \non 
\eea

Other choices for $\langle Y\rangle$ with
$n_1 \neq 0 \neq n_2$ would not change the tree level potential
energy of the vacua, so there is a landscape of vacua at
classical level. This is wiped out by 1-loop corrections.
We can give more details about the 
computation in the case where $n_1\neq 0$ and how
we excluded the possibility of a landscape at quantum level.
We parametrize the fluctuations around the non supersymmetric vacua
in the case of non trivial vev for the adjoint field
\be
\label{para1}
q=\left ( \begin{array}{c}
k e^{\theta}+\xi_1\\
\phi_1\\
\end{array}\right) 
\qquad 
\tilde q^{T} =\left ( 
\begin{array}{c}
k e^{-\theta}+ \xi_2\\
\phi_2\\
\end{array}\right)
\qquad
\langle Y \rangle= 
\left( \begin{array}{c  c}
y_1&0 \\
0&y_2 \\
\end{array} \right)+\delta Y
\ee
\be
\label{para2}
M_1=\left ( \begin{array}{c c}
p_1^A+\xi_3&\phi_3\\
\phi_4&p_1^B+\xi_4\\
\end{array}\right) 
\qquad
M_2=\left ( \begin{array}{c c}
p_2^A+\xi_5&\phi_5\\
\phi_6&\mc{X}\\
\end{array}\right)
\ee 
The resulting superpotential for the sector affected
by the supersymmetry breaking (the $\f_i$ chiral fields)
is
\bea
W&=&h_2 \left( \mathcal{X} \f_1 \f_2-m_2^2 \mc{X} \right)+h_2 m_2 \left( e^{\theta} \f_2 \f_5
+ e^{-\theta} \f_1 \f_6 \right) + \non \\
\label{orafe2} 
&&+ (h_1+h_3 y_i)m_2\left( e^{\theta} \f_2 \f_3+ e^{-\theta} \f_1 \f_4 \right) + 2 m_3 \f_3 \f_4 
+\frac{h_1 m_1^2}{2 m_3}(h_1+h_3 y_i) \,  \f_1 \f_2
\eea
where $i=1,2$.
Exactly we have $n_1$ copies of (\ref{orafe2}) with $i=1$ and $\tilde N-n_1$ copies with 
$i=2$.
The fields appearing in (\ref{orafe2}) are the only ones which
contribute to the one loop potential.

Comparing with the one in \cite{Amariti}, 
we observe that having $n_1 \neq 0$ contributes only in 
a shift in the $\zeta$ and $\rho$ parameters.
We can then compute
the 1-loop quantum corrections 
to the scalar potential $V^{1-loop}(n_1)$,
which depends non trivially on $n_1$ through $y_i(n_1)$.
This contribution is minimized when $n_1=0$, i.e.
 $\langle Y \rangle=0$, implying 
that this is the lowest energy vacuum \cite{Amariti}.

\section{Decay to the supersymmetric vacuum}
Supersymmetry is restored when the $SU(\tilde N)$ symmetry is gauged
via non perturbative effects \cite{Affleck}, 
away from the metastable vacuum
in the field space.
The non supersymmetric vacuum 
is then a metastable state of the theory
which decays into a supersymmetric one.
In \cite{Amariti} we find a supersymmetric vacuum in
the large field region for the meson $M_2$ 
where 
\be
\label{M_2}
\langle h_2 M_2 \rangle=\tilde \Lambda \epsilon^{\frac{\tilde N}{N_f-\tilde N}} \xi^{\frac{\tilde N}{N_f-\tilde N}}\mathbf{1}_{N_f}
=m_2 \left( \frac{1}{\epsilon}\right)^{\frac{N_f-2 \tilde N}{N_f-\tilde N}} \xi^{\frac{\tilde N}{N_f-\tilde N}} \mathbf{1}_{N_f} \, ,
\qquad 
\epsilon=\frac{m_2}{\tilde \Lambda} \quad \xi=\frac{m_2}{m_Y} \,.
\ee
$\epsilon$ is a dimensionless parameter which can be made parametrically
small sending the Landau pole $\tilde \Lambda$ to infinity.
$\xi$ is a dimensionless finite parameter which 
does not spoil the estimation of the supersymmetric vacuum in the sensible
range $\epsilon<\frac{1}{\xi}$.
All the exponents appearing in (\ref{M_2}) are positive in the window (\ref{range}). 

We observe that in the small $\epsilon$ limit the vev $\langle h_2 M_2 \rangle$ is larger than the
 mass scale $m_2$ of the magnetic theory 
but much smaller than
the scale $\tilde \Lambda$
\be
m_2 \ll \langle h_2 M_2 \rangle \ll \tilde \Lambda.
\ee
making the evaluation of the supersymmetric vacuum reliable.

We now make a qualitative evaluation of the decay rate of the metastable vacuum. 
At semi classical level the decay probability is proportional to
$e^{-S_B}$ where $S_B$ is the
bounce action from the non supersymmetric vacuum to a supersymmetric one.
We obtain as the decay rate \cite{Amariti}
\be
S\sim \left(\left( \frac{1}{\epsilon}\right)^{\frac{N_f-2 \tilde N}{N_f-\tilde N}} \xi^{\frac{\tilde N}{N_f-\tilde N}} \right)^4
\sim \left( \frac{1}{\epsilon}\right)^{4 \frac{N_f-2 \tilde N}{N_f-\tilde N}}
\ee
This rate 
can be made parametrically large sending to zero the dimensionless
ratio $\epsilon$ (i.e. sending $\tilde \Lambda \to \infty$)
 since the exponent $ \left( 4 \frac{2 \tilde N-N_f}{\tilde N- N_f}\right)$ is always
positive in the window (\ref{range}).

\section{Massless quarks}\label{masszero}
The non supersymmetric meta-stable vacua survive  
the limit $m_Q \to 0$. Indeed this limit
corresponds in the magnetic description to send to zero the linear term
for $M_1$, i.e. $m_1 \to 0$, and 
all the results are smooth in this limit.

At classical level 
there are only small differences in the field vacuum expectation
values.
At quantum level this limit set the parameter $\zeta$ to zero, 
but the qualitative behaviour of the
1-loop corrections is the same: the classical flat directions
are still lifted. 

Finally the computation of the supersymmetric vacuum for $m_1=0$ is
even more straightforward, giving a vanishing vacuum expectation values 
for the meson $M_1$.
The lifetime estimation of the non-supersymmetric vacuum is not affected
by this limit, and it is still parametrically large. \\
Setting $m_1=0$ the model become 
more similar to the one studied in \cite{ooguri}.

\section{R-symmetry and gauge mediation}\label{massagaugino}
We are interested in direct gauge mediated supersymmetry breaking.
In this framework the gauge group of the SM has to be
embedded 
into a flavour group of
the dynamical sector.
The gauge sector of the SM  
directly couples to the supersymmetry breaking dynamics
and a natural question for model building is whether the gauginos
of the MSSM acquire masses. 

We can embed the SM gauge group 
into the subgroups of
the flavour symmetry $SU(2 N_f-N_c)$ or $SU(N_c-N_f)$
provide $(2 N_f-N_c>5)$ or $(N_c-N_f>5)$, respectively.
As in \cite{ooguri3} we can compute the
beta function coefficient $b_{SU(3)}$ at different 
renormalization scales and 
we conclude that in order to avoid Landau pole problems
the embedding should be done in $SU(2 N_f-N_c)$.

The full model has no 
R-symmetry,
and, unlike \cite{Seiberg,ooguri}, no accidental R-symmetry
arises at the non-supersymmetric meta-stable vacuum,
and hence a gaugino mass generation is not forbidden \cite{Seiberg}.
Moreover the absence of R-symmetry implies that the 
non supersymmetric minimum is not at the origin of the moduli
space, i.e. $\langle \mathcal{X} \rangle \neq 0$.

The $R$-breaking terms are the quadratic massive terms  
$\phi_1 \phi_2$ and 
$\phi_3 \phi_4$ in (\ref{orafe2}). The first one can be eliminated
shifting the field $\mathcal{X}$. The second one cannot
be eliminated rearranging the fields. If
the mass $m_3$ is larger than the supersymmetry breaking  
scale, $\phi_3$ and $\phi_4$ could be integrated
out, supersymmetrically, recovering an accidental $R$-symmetry:
this, however, is not our range of parameters.

We analyze the dynamics at the meta-stable vacuum
where the breaking of supersymmetry
generates a gaugino mass proportional to
the breaking scale $F_{\mathcal{X}}$.
Contribution to this mass comes from the
superpotential\footnote{For simplicity
we consider only one copy of the chiral superpotentials.} of the messengers $\f_i$
\be
\label{supgaugino}
W \supset h_2 \left( \mathcal{X} \f_1 \f_2 \right)+h_2 m_2 \left(\f_2 \f_5
+ \f_1 \f_6 \right) + h_1 m_2\left( \f_2 \f_3+ \f_1 \f_4 \right) + 2 m_3 \f_3 \f_4 
+\frac{h_1^2 m_1^2}{2 m_3} \,  \f_1 \f_2
\ee
which, in a matrix notation, reads
\be
\left( \begin{array}{ccc}
\f_1&\f_3&\f_5
\end{array}\right) 
\mathcal{M}
\left( 
\begin{array}{c}
\f_2\\
\f_4\\
\f_6
\end{array}\right) 
\ee
where $\mathcal{M}$ is a mass matrix for the messenger fields
\be
\label{massgaug}
\mathcal{M}=\left ( \begin{array}{c c c}
h_2 \langle \mathcal{X}\rangle+\frac{h_1^2 m_1^2}{2 m_3}&h_1 m_2&h_2 m_2\\
h_1 m_2&2 m_3&0\\
h_2 m_2&0&0\\
\end{array}\right)\equiv \, h_2 m_2
\left ( \begin{array}{c c c}
\frac{\langle \mathcal{X}\rangle}{m_2}+ \zeta & \rho &1\\
\rho &\eta&0\\
1&0&0\\
\end{array}\right)
\ee
This matrix does not generate a gaugino mass
at one loop at first order in $F_{\mathcal{X}}$
as in \cite{Izawa}.
However at the third order in 
$F_{\mathcal{X}}$,
the gaugino mass arises
as in \cite{Csaki,Izawa}.
This contribution is not negligible when 
$\frac{F_{\mathcal{X}}}{h_2^2 m_2^2} \sim 1$, 
which is admitted in
our range of parameters.

Diagonalization of (\ref{massgaug}) and use
of the general formula in \cite{Martins} for the
computation of the 1 loop diagrams contributing to
the gaugino mass $m_{\lambda}$ lead to
\be
\label{massgaugino}
m_{\lambda} \sim \frac{F^3_{\mathcal{X}}}{(h_2 m_2)^5} \left[
\frac{1}{4}\left( \frac{\langle \mathcal{X}\rangle}{m_2}+ \zeta \right)
+ \rho^2 \eta  \right]
\ee
The coefficient of $F^3_{\mathcal{X}}$ in (\ref{massgaugino}) is evaluated at the 
third order in 
the adimensional small parameters ($\rho,\eta,\zeta$):
indeed by direct inspection we find that also
the term $\left( \frac{\langle \mathcal{X}\rangle}{m_2}+ \zeta \right)$
gives at least third order contributions in ($\eta,\rho,\zeta$).

\section*{Acknowledgments}
We would like to acknowledge K. Intriligator, R.Rattazzi, N.Seiberg and 
A.M.Uranga for stimulating discussions.
This work has been supported in part by INFN, by PRIN prot.2005024045-002 
and the European Commission RTN program MRTN-CT-2004-005104.

\end{document}